\documentclass[twocolumn,prl,showpacs,floatfix,preprintnumbers,amsmath, amssymb,superscriptaddress]{revtex4}

\usepackage{amsmath}
\usepackage{graphicx}
\usepackage{subfigure}
\usepackage{psfrag}
\usepackage[centerlast]{caption}
\usepackage[dvips]{color}

\usepackage{graphicx}
\usepackage{dcolumn}
\usepackage{bm}
\usepackage{units}

\newcommand{\be}{\begin{equation}}
\newcommand{\ee}{\end{equation}}
\newcommand{\rv}{\boldsymbol{r}}
\newcommand{\vv}{\boldsymbol{v}}

\pacs{47.32.-y,03.75.Lm,03.75.Kk}

\begin{document}

\title{Vortex-Induced Phase Slip Dissipation in a Toroidal \\ Bose-Einstein Condensate  
Flowing Through a Barrier}

\author{F. Piazza}\affiliation{CNR-INFM BEC Center and Dipartimento di Fisica, Universit\`a di Trento, I-38050 Povo, Italy}
\author{L. A. Collins}\affiliation{Theoretical Division, Mail Stop
B214, Los Alamos National Laboratory, Los Alamos, New Mexico 87545}
\author{A. Smerzi}\affiliation{CNR-INFM BEC Center and Dipartimento di Fisica, Universit\`a di Trento, I-38050 Povo, Italy}

\date{\today}
                      
 \begin{abstract}
We study superfluid dissipation due to phase slips 
for a BEC flowing through a repulsive barrier inside a torus.
The barrier is adiabatically raised across the annulus while the condensate flows 
with a finite quantized angular momentum.
At a critical height, a vortex moves from the
inner region and reaches the barrier
to eventually circulate around the annulus.
At a higher critical height, an anti-vortex also enters into the torus from the outer region.
Both vortex and anti-vortex
decrease the total angular momentum by leaving behind a $2\pi$ phase slip. 
When they collide and annihilate or orbit along the same loop,
the condensate suffers a global $2\pi$ phase slip, and the total angular momentum
decreases by one quantum. 
In hydrodynamic regime, the instability sets in when the local superfluid velocity equals the sound speed inside the barrier region. 

\end{abstract}

\maketitle

{\it Introduction}.
Flow dynamics through a constriction can reveal essential aspects of
superfluidity. A central feature observed long ago with superfluid $\mathrm{^4He}$ 
currents through an orifice \cite{Avenel_1985} is the occurrence
of single $2\pi$ phase slips, which collectively decrease the fluid velocity
by a quantized amount. 
More recently, the transition from phase slips to the Josephson regime has been observed 
by increasing the helium healing length with respect to the size of the orifice
\cite{Packard_2006}.

Common belief associates
phase slips with the nucleation of vortices
transversally crossing the constriction \cite{Anderson_1966}. 
This mechanism has been invoked to explain the dissipation of the superfluid helium flow, which occurs 
at critical velocities much lower than predicted by the Landau criterium.
The microscopic mechanism of the onset of the instability 
and its dynamical evolution, however, are still not completely
understood \cite{Varoquaux_2006}. 

The superflow dynamics of
a dilute Bose-Einstein condensate (BEC) gas can shed new light on the physics of phase slips. 
While in quantum liquids constrictions are
made by single or multiple orifices, in BECs they can
be created by a laser beam generating a repulsive barrier for the atoms, or by an offset of the central hole in toroidal geometries \cite{Phillips_2007}.  
Broadly speaking, similar configurations allow for the observation
of macroscopic phase coherence effects and can
lead to a range of important technologies.
While superconducting Josephson junctions are already employed in 
sensors and detectors, their superfluid counterparts can realize
ultrasensitive gyroscopes to detect rotations \cite{Packard_2006}.
For instance, a
toroidally shaped superfluid weak link provides the
building block of a d.c.-SQUID, which is 
most promising sensing device based on superfluid interference.

A distinctive feature of quantum gases rests with the possibility of
experimentally interrogate the response of the
system in a wide variety of traps and dynamical configurations.
Moreover, even if the BEC is described by a local Gross-Pitaevskii Eq.(\ref{GP}), 
(as in most cases where dipolar interactions can be neglected) 
and therefore lacks the rotonic part of the helium spectrum, 
its nonlinearity appears to be the only crucial ingredient needed
to reveal the microscopic mechanisms underlying the vortex-induced
phase-slips. 
Superfluidity of a BEC confined in a
torus, in  absence of barriers,  has been first experimentally observed at NIST \cite{Phillips_2007}. The BEC was initially stirred by transfer of 
quantized orbital angular momentum from a
Laguerre-Gaussian beam and the rotation remained stable up to 10 seconds
in the 
multiply connected trap. 
The metastability of a ring-shaped superflow due to centrifugal forces has been observed in \cite{Cornell_2003}.
The superfluid critical velocity in a harmonically trapped BEC
swept by a laser beam has been observed experimentally in
\cite{Ketterle_1999} and associated with the creation of vortex phase singularities in \cite{Inouye2001} while solitons were observed in the one dimensional geometry of \cite{Engels_2007} .
Such problems have been object of a large theoretical study mainly based on numerical simulations of the GPE \cite{Rica_1992}.

\begin{figure}[]
\hspace{-0.0in}
\includegraphics[scale=.7]{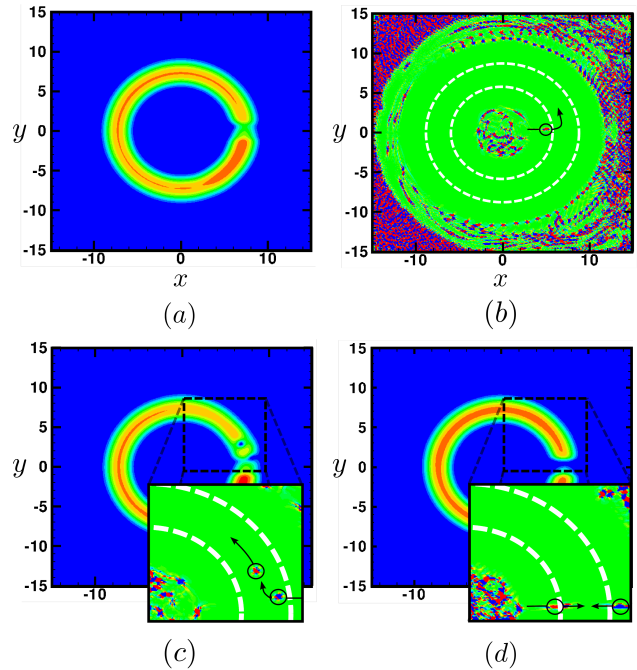}
\vspace{-0.25in}
\caption{\small{(Color online)  (a), (b) and (c) $t_r=10$, $L_z/N=8$ and $V_s=0.34\ \mu$.
    (a) $t=7.6$. Density contour plot with no visible vortex core. (b) $t=7.6$. The $z$ component of
    the vorticity field $\nu(\rv)$. The
    white dashed lines indicates the TF radii of the cloud. The
    encircled dot corresponds to a vortex about to
    enter the annulus from the inner edge. (c) $t=11.6$. A vortex 
    circulates along the annulus while the vorticity (inset)
    shows an anti-vortex about to enter. (d) $t_r=10$, $L_z/N=2$, $V_s=0.61\ \mu$. Vortex anti-vortex annihilation. }} \label{f1}
\vspace{-0.2in}
\end{figure}

In this manuscript, we theoretically study the dynamics of a BEC flowing inside a
toroidal trap at zero temperature and in the presence of a repulsive barrier. Similar qualitative results are observed when, rather than by a repulsive barrier, the constriction is created by an offset in the position of the central hole of the torus.
As initial condition, we consider a superfluid state with a finite orbital angular momentum
in the cylindrically symmetric torus. The
critical regime is reached by adiabatically raising the standing
repulsive barrier. The dissipation takes place through phase slips created by
singly-quantized vortex lines crossing
the flow.  
We found two different critical barrier heights.
At the smallest critical height, 
a singly-quantized vortex moves radially along a straight path from the center of the
torus and enters the annulus (Fig. \ref{f1}(a)-(b)), leaving behind 
a $2 \pi$ phase
slip. Eventually, it keeps circulating with the
background flow without crossing completely the torus so that it decreases the total angular momentum only by a fraction
of unity. At the highest critical height, 
a singly-quantized anti-vortex enters the torus from the outward low density region of the system. 
The ensuing vortex dynamics depends on the velocity asymmetry between the inner and the outer edge of the annulus as well as on the final barrier height and ramping time.
For instance, a vortex and an anti-vortex can just circulate on separate orbits (Fig. \ref{f1}(c)) or can collide along a radial trajectory and annihilate (Fig.~\ref{f1}(d)). When they orbit on the same loop or annihilate, the system undergoes a global $2\pi$ phase slip, with the decrease of one unit of total angular momentum. In general, the BEC flow can be
stabilized after the penetration of a few vortices.

In hydrodynamic regime, we find that the instability towards vortex penetration occurs when the local superfluid velocity equals the sound speed. This happens inside the barrier region and close to the edges of the cloud.
We have studied the above scenario in 2D and 3D numerical simulations of the dynamical GPE.
In the 3D analysis, we have employed the experimental
parameters of the toroidal trap created at NIST \cite{Phillips_pc}.
The experimental investigation of the system proposed here can provide the 
first direct observation of interconnection and dynamical evolution of
vortices and phase slips in superfluid systems. 

{\it Phase-slips and vortices.}
We numerically solve the time dependent GPE
\begin{equation}\label{GP}
i \hbar \frac{\partial\psi(\rv,t)}{\partial t}=\left[-\frac{\hbar^2 \nabla^2}{2 m}+V_{t}(\rv)+V_{b}(\rv,t)+g|\psi|^2\right]\psi(\rv,t)
\end{equation}
where $g$ is proportional to the inter-particle scattering length.
In the following, we first consider an effective 2D Cartesian geometry \cite{2Dgrid} and eventually extend the analysis to the
3D configuration. 
The trapping potential $V_t(\rv)=V_{h}(\rv)+V_c(\rv)$ is
made by an harmonic confinement $V_h(\rv)=\hbar\omega_{\perp}(x^2+y^2)/2d_{\perp}^2$ plus a 
gaussian core $V_{c}(\rv)=V_0\exp[-(x^2+y^2)/\sigma_c^2]$ creating a
hole in the trap center (in what follows we will express quantities in
trap units of time $\omega_{\perp}^{-1}$ and length $d_{\perp}$).
As an initial condition, we consider the numerical
ground state obtained with $V_b=0$ and  transfer by linear phase imprinting (in 3D calculations a Laguerre-Gaussian beam is implemented \cite{Phillips_pc}) 
a total angular momentum $L_z=N l$, with $N$ the total number of
particles and $l$ integer. The transferred angular momentum is low enough to have flow velocities in the torus region much smaller than the sound
speed.
Over each loop of radius $r=\sqrt{x^2+y^2}$
the circulation is $C=2\pi l$ 
and the modulus of the fluid velocity, $v(r)=C/2\pi r$, is constant
and directed
along the tangent of the same loop. In
principle, these $l$ quanta of circulation can be carried by a single
multiply-quantized macro-vortex \cite{Stringari_2006}, which hovewer breaks up into singly-quantized vortices still
confined within the central hole \cite{macrovortex}. 
In our simulations, 
as soon as a finite angular momentum is transferred to the condensate,
the vorticity field component perpendicular to the $x-y$ plane
$\nu(\rv,t)=(\nabla\times\vv(\rv,t))\cdot\hat{z}$ shows a ``sea'' of positive and negative vorticity
spots, that is, a mesh of vortices and anti-vortices,
Fig.~\ref{f1} (b). 
This happens in two regions of very
low density, close to the center and in the 
space surrounding the torus
\cite{vortex_mesh}. 

After angular momentum is transferred to the cloud, the barrier potential $V_b(\rv,t)$ is slowly ramped
up over a time $t_r$ to a final heigth $V_s$. We use a repulsive well with widths $w_x$
centered at the maximum density and $w_y$ centred at $y=0$ \cite{barrier}. We always choose $w_x>d$, where $d\equiv R_2-R_1$ is the width of the annulus. 
Initially, the density and velocity field adapt to the
presence of the barrier, and the flow shows no sign of
dissipation. In the barrier region, where the density is depleted, the flow velocity increases mainly at the edges of the annulus.
By examining the vorticity, we observe that the two vortex seas are
strongly fluctuating, with vortices and anti-vortices trying to escape
but being pushed back by zones of higher density. However,
when the barrier reaches a critical height $V_{c1}$, 
a vortex from the inner sea can successfully escape and enter the annulus.
As shown in Fig.~\ref{f1} (a) and (b) \cite{nota10}, at $V_{c1}$ the flow can no longer sustain a stationary configuration and becomes unstable.
In Fig.~\ref{f1} (a), we observe the depletion of
the density but not a visible vortex core. However, if we inspect
the the vorticity field plotted in Fig.~\ref{f1} (b), we
clearly see an isolated red spot, corresponding to a positive vorticity, moving radially from the center of the torus towards the
higher density region, indicating the presence of the core of a singly-quantized vortex \cite{ghost}. 

The above scenario for vortex nucleation in a multiply
connected geometry confirms that a persistent
flow in such a configuration is possible because of the pinning of the vorticity in 
the low density regions near the center and outside of the torus. The pinning is
due to the effective energy barrier felt from a vortex core when trying to move towards a region of
much higher density \cite{Abraham_2001}. 
The effective energy barrier arises from the nonlinearity of the GPE.  
The obstacle raised across the annulus serves to unpin singly-quantized vortices by steadily decreasing the density during the ramping process, up to suppression of the effective energy barrier.
The density depletion occurs on a radial stripe 
and makes way for the vortex moving outwards along a straight line connecting the center of the vortex with 
the barrier. This as well happens for the anti-vortex moving inwards at a larger height of the repulsive barrier, see below.  

In the hydrodynamic regime, when $\xi\ll d, w_x, w_y$ and $V_s\ll\mu$, we observe the instability towards vortex penetration when the local superfluid velocity reaches the sound speed \cite{criterion}. This critical condition is first met inside the barrier region, at the Thomas-Fermi radius of the cloud. The sound speed, is calculated at the maximum of the repulsive well (at $y=0$ in our case) with the density integrated along the radial direction \cite{analytical}. 
The parameters of Fig.~\ref{f1}, however, have been chosen such that the system is outside the hydrodynamic regime, in order to emphasize the generality of the presented vortex dynamics phenomenology.
\begin{figure}[]
\hspace{-.in}
\includegraphics[scale=.4]{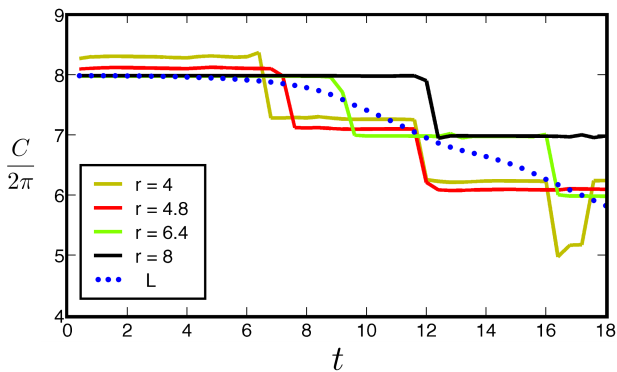}
\caption{\small{(Color online)
Circulation (solid lines) for loops with different radii and total angular
momentum (dots) as a function of time. The parameters 
are the same as in Fig.~\ref{f1}(a).
The $2\pi$ drops in the circulation at $r=4,4.8,6.4$
are due to a singly-quantized vortex moving
outwards from the center. The drop at $r=8$ and $t\sim 12$ is due
to the passage of an anti-vortex entering the annulus from the outer edge. 
The oscillation in the circulation at $r=4$ and $t\sim 16-17$
is due to a double crossing of a vortex trying to escape the
inner region.}} \label{f2}
\vspace{-0.in}
\end{figure}

The passage of a vortex core between two points causes a $2\pi$ slip in
the phase difference between them \cite{Anderson_1966}. In
Fig.~\ref{f2}, we observe $2\pi$ sharp drops in the circulation $C$ on a
given loop of radius $r$ at the moment the vortex core crosses
it. 
Moreover, at
small radii, close to the inner sea of vortices,
the circulation shows spikes at which it decreases by $2\pi$, then quickly
goes back to its previous value. These are associated with a vortex
moving out of the sea but being pushed back by a region
of high density located slightly outwards, as discussed above. 

Due to phase slippage, the
angular momentum is reduced, and eventually the system becomes stable again
after a finite number of spawned vortices. The circulation is lowered
by a few
quanta, and the fluid velocity on vortex-crossed loops is brought back
below the critical value. If the ramping is stopped at $V_{c1}$, only
the inner edge of the annulus is unstable since its fluid velocity is
larger ($v(r)\propto l/r$) 
. In this case, vortices do not cross
completely the torus and move 
on stable circular orbits \cite{berloff} . 

However, when the barrier reaches a second critical height $V_{c2}>V_{c1}$ the outer part of the annulus becomes also
unstable. Anti-vortices then enter from outwards while vortices enter
the inner edge, as previously discussed. Anti-vortices move radially inwards
and contribute to stabilize the outer part by phase slips. Indeed, an anti-vortex crossing a loop makes the circulation drop as a vortex crossing the opposite way.  
In Fig.~\ref{f1}(c)  we see a vortex already circulating inside the
high density region of the annulus while an anti-vortex begins to
enter. 
The separation between $V_{c1}$ and $V_{c2}$ is proportional to the velocity difference $\Delta v=l(R_1-R_2)/(R_1R_2)$ between the two edges. 
In general, depending on $\Delta v$, the dynamics at barrier heights larger than $V_{c2}$ can vary. 
For instance, at lower angular momenta $\Delta v$ becomes smaller, and a vortex and
an anti-vortex enter the annulus almost simoultaneously. They can then collide and annihilate, as shown in Fig.~\ref{f1}(d). 
When a vortex and an anti-vortex annihilate or separately orbit on the same loop, the system undergoes a global $2\pi$ phase
slip, and the total angular momentum is decreased by one unit. 

We extend our 2D calculations into a 3D
configuration \cite{3Dgrid}. The parameters of the 
toroidal trap are those
employed experimentally at NIST \cite{Phillips_pc}. 
We add a repulsive well
potential \cite{barrier} whose
shape, however, is not crucial in determining the qualitative features of
the dissipation as long as $w_x$ is larger than the width of the annulus. Since the healing length 
is of the order of the harmonic lenght along z we found, as expected, 
that the nucleation of singly-quantized vortex lines 
and their dynamics resemble those observed
in 2D calculations. In particular, we have two critical
values for the barrier height $V_{c1}$ and $V_{c2}$ connected respectively with the nucleation of vortices or both vortices and anti-vortices.

{\it Conclusions.}
We have studied the superfluid dynamics of a dilute Bose-Einstein condensate confined in a toroidal trap
in presence of a repulsive barrier.
With a finite
initial angular momentum, we observed two
critical values of the barrier height for the onset of phase slips dissipation: a lower one corresponding to vortices entering the
annulus from the center of the torus, and a higher one related to both
vortices and anti-vortices, the latter entering the outer edge of the
annulus. 
We have performed 3D simulations with the NIST toroidal trap parameters, where
the above scenario could be experimentally observed when a standing repulsive
barrier is raised across the BEC supefluid flow. Since
supercurrents have recently been observed in absence of the barrier, we believe that the
experimental confirmation of our results is at hand. 
Vortices can be directly observed with BECs, and it is therefore possible to experimentally
characterize their role in phase-slips-induced dissipation in superfluid systems. 

{\it Acknowledgements.}
We would like to thank B. Schneider, F. Dalfovo, L. Pitaevskii, and S. Stringari for helpful discussions and Dr. S. Hu for assistance with the 3D GPE program.
 We acknowledge useful exchanges with W. Phillips, S. Muniz, A. Ramanathan, K. Helmerson, and P. Clade. The Los Alamos National Laboratory is operated by Los Alamos National Security, LLC for the National Nuclear Security Administration of the U.S. Department of Energy under Contract No.~DE-AC52-06NA25396.


\begin{thebibliography}{99}
\bibitem{vortex_mesh} 
  We can regard the presence of the ``vortex sea'' as due to numerical noise acting
  inside very low density regions, possibly triggering a dynamical
  instability of GPE.
\bibitem{barrier} The barrier potential is $V_b(\rv,t)=f(t)  {{V_s }} V_{bx}(x)V_{by}(y)/4$, with $f(t)=t/t_r$ ($f(t)=1$ for
$t>t_r$) and $V_{bx}=\tanh(\frac{x-R_x+x_0}{b_s})+\tanh(\frac{-x+R_x+x_0}{b_s})$.
Here $R_x$ is the $x$-shift of the center of the barrier while its width is $w_x\sim 2 x_0$. $V_{by}$ has the same shape as $V_{bx}$ but 
with $R_y=0$. The final height of the barrier is
$V_s$ as long as $x_0,y_0\gg b_s$.
\bibitem{nota10}  Before raising the barrier, the flow velocity
at the maximum density, located at $r_m=7.2$, is
$v(r_m)=0.67\ c_s$, where $c_s$ is Thomas-Fermi (TF) estimate of the sound speed in the ground state. The TF value of the healing length at $r_m$ is $\xi\sim
0.28$.  The TF width of the annulus is $d=3.59$. In this case, the barrier widths are $w_x\sim 4$ and $w_y\sim 2$. The observed critical barrier heights
are $V_{c1}\sim 0.14\ \mu$ and $V_{c2}\sim 0.24\ \mu$.
\end{thebibliography}
\end{document}